\documentclass[12pt]{article}
\usepackage{graphics}
\usepackage{epsfig}
\usepackage{subfigure}
\usepackage{graphics}
\usepackage{amsmath}
\usepackage{amsfonts}
\usepackage{amssymb}
\usepackage{url}
\usepackage{hyperref}
\newcommand{\MSUSY}{\left( \frac{F_{\phi}}{16\pi^2} \right)}

\newcommand{\be}{\begin{equation}}
\newcommand{\ee}{\end{equation}}
\newcommand{\bea}{\begin{eqnarray}}
\newcommand{\eea}{\end{eqnarray}}

\newcommand{\GeV}{~\mathrm{GeV}}

\newcommand{\zp}{Z^{\prime}}

\def\nn{\hspace{2mm}}
\def\sss{\scriptscriptstyle}
\def\GeV{\mbox{\rm GeV}}

\def\abs#1{\left| #1\right|} 

\begin{document}
%%%%%%%%%%%%%%%%%%%%%%%%%%%%
\begin{titlepage}
%%%%% PREPRINT NUMBERS %%%%%%
\begin{flushright}
KEK-TH-1265
\end{flushright}
%%%%%%%%%%%%%%%%%%%%%%%%%%%%%%
\vspace{4\baselineskip}
%%%%%%%%%%%%%%%%%%% TITLE %%%%%%%%%%%%%%%%%%
\begin{center}
{\Large\bf 
$B-L$ assisted Anomaly Mediation and\\
the radiative $B-L$ symmetry breaking
}
\end{center}
%%%%%%%%%%%%%%%% AUTHORS %%%%%%%%%%%%%%%%%%%%%%%
\vspace{1cm}
\begin{center}
{\large
Tatsuru Kikuchi$^{a}$
\footnote{\tt E-mail:tatsuru@post.kek.jp}
and Takayuki Kubo$^{a,b}$
\footnote{\tt E-mail:kubotaka@post.kek.jp}
}
\end{center}
%%%%%%%%%%%%%%%%%%%%%%% AFFILIATION %%%%%%%%%%%%
\vspace{0.2cm}
\begin{center}
${}^{a}$ 
{\small \it Theory Division, KEK,
Oho 1-1, Tsukuba, Ibaraki, 305-0801, Japan}\\
${}^{b}$ 
{\small \it The Graduate University for Advanced Studies,
Oho 1-1, Tsukuba, Ibaraki, 305-0801, Japan}\\
\medskip
\vskip 5mm
\end{center}
\vskip 5mm
\begin{abstract}
Anomaly mediated supersymmetry breaking implemented in the minimal supersymmetric standard model (MSSM) 
is known to suffer from the tachyonic slepton problem leading to breakdown of electric charge conservation.
We show however that when MSSM is extended to explain small neutrino masses by gauging the B-L symmetry, 
the slepton masses can be positive due to the $Z^\prime$ mediation contributions.
We obtain various soft supersymmetry breaking mass spectra, which are different
from those obtained in the conventional anomaly mediation scenario.
Then there would be a distinct signature of this scenario at the LHC. 
\end{abstract}
\end{titlepage}
%%%%%%%%%%%%%%%%%%%%%%%%%%%%%%%%%%%%%%%%%%%%%%%%
\section{Introduction}
%%%%%%%%%%%%%%%%%%%%%%%%%%%%%%%%%%%%%%%%%%%%%%%%
Supersymmetry (SUSY) extension is one of the most promising way 
 to solve the gauge hierarchy problem in the standard model (SM) \cite{SUSY}.
Since any superpartners have not been observed 
 in current experiments, SUSY should be broken at low energies. 
Furthermore, soft SUSY breaking terms are severely constrained 
 to be almost flavor blind and CP invariant. 
Thus, the SUSY breaking has to be mediated to the visible sector 
not to induce too large CP and flavor violation effects.
Some mechanisms to achieve such SUSY breaking mediation  
 have been proposed \cite{Luty:2005sn}. 

The anomaly mediated supersymmetry breaking (AMSB) scenario
\cite{AMSB1, AMSB2, AMSB3} is one of the most attractive scenario 
 due to its flavor-blindness and ultraviolet (UV) insensitivity 
 for the resultant soft SUSY breaking terms.
The pattern of SUSY breaking does not depend
at all on physics at higher energy scales.
On the eve of the Large Hadron Collider (LHC) operation at CERN, 
which start this year, there are several studies in the aspects of 
collider physics to discriminate the AMSB scenario from 
the other SUSY breaking mediation scenarios \cite{Barr:2002ex, Datta:2002vy, Asai:2007sw}.
Despite the appeal of the AMSB,
the original version of the AMSB is excluded because of its high predictivity.
The slepton squared masses become negative at the weak scale,
and hence the theory would break U(1)$_{\rm em}$.
There have been many attempts to solve this problem
by incorporating additional positive contributions
to slepton squared masses at tree level \cite{AMSB1, tree1, tree2, Kitano:2004zd}
or at quantum level \cite{PR, quantum}.

An important thing to realize at this point is that MSSM is not a complete theory of low energy
particle physics and needs extension to explain the small neutrino masses observed in experiments.
The relevant question then is whether MSSM extended to include new physics that explains small
neutrino masses will cure the tachyonic slepton mass pathology of AMSB.
One of the simplest extensions of MSSM which provide natural explanation of small neutrino
masses is to extend the gauge symmetry of MSSM to $SU(3)_c \times SU(2)_L \times U(1)_Y \times U(1)_{B-L}$ 
which naturally introduces three right-handed neutrinos into the theory in order for the anomaly cancellation.
Once we incorporate the U(1)$_{B-L}$ gauge symmetry in SUSY models, the U(1)$_{B-L}$ gaugino $\tilde{Z}_{B-L}$ appears,
and it can mediate the SUSY breaking \cite{KK} (the $Z^\prime$ mediated SUSY breaking \cite{ZMSB1, ZMSB2}) 
\footnote{The similar idea has also been suggested in \cite{Mohapatra:1997kv, Dobrescu:1997qc}.}.
The prensent papaer focuses on an alternative approach to avoid the tachyonic slepton problem, where we use 
the $Z^\prime$ mediated SUSY breaking \cite{ZMSB1, ZMSB2}.

This paper is organized as follows. In Sec. 2 and Sec. 3, we give a brief review of the anomaly mediation
and the $Z^\prime$ mediated SUSY breaking, respectively.
In Sec. 4, we combine these two scenarios and examine the numerical evaluations
to give a sample mass spectra.
Sec. 5 is devoted to summary and discussion.

%%%%%%%%%%%%%%%%%%%%%%%%%%%%%%%%%%%%%%%%%%%%%%%%%%%%%%%
%%%%%%%%%%%%%%%%%%%%%%%%%%%%%%%%%%%%%%%%%%%%%%%%%%%%%%%
\section{Anomaly Mediation in the $B-L$ extended MSSM}
%%%%%%%%%%%%%%%%%%%%%%%%%%%%%%%%%%%%%%%%%%%%%%%%%%%%%%%
%%%%%%%%%%%%%%%%%%%%%%%%%%%%%%%%%%%%%%%%%%%%%%%%%%%%%%%
In this section, we work out in the superconformal framework 
 of supergravity \cite{Kaku:1978nz}, 
and we explain the anomaly mediation scenario in the $B-L$ extended MSSM.

In the superconformal framework of supergravity, 
the basic Lagrangian is given by 
\bea
 {\cal L}_{\rm SUGRA} &=& - 3 \int d^4 \theta \;\phi^\dag \phi \; e^{-K/3}  
  + \int d^2 \theta \;\phi^3 W + h.c. \; , 
\eea
where $\phi=1+ \theta^2 F_{\phi}$ is the compensating multiplet,
$K$ is the K\"ahler potential in the conformal frame, and $W$ is the superpotential.

As for the gauge sector in the MSSM, the kinetic term is of the form, 
\bea
{\cal L}_{\rm gauge} = \frac{1}{4} \int d^2 \theta \;
 \tau_a\left(\frac{\mu_R}{\Lambda \phi} \right)  {\cal W}^{a \alpha} {\cal W}^a_\alpha \; . 
\label{gaugeK}
\eea
At the classical level, the compensator $\phi$ does not appear
in the gauge kinetic term as the gauge chiral superfield ${\cal W}^{a \alpha}$
has a chiral weight $\frac{3}{2}$. 
It turns out that the dependence of $\phi$ comes out radiatively
through the cutoff scale $\Lambda$ ($\mu_R$ is the renormalization scale).
In the above setup, non-zero $F_\phi$ induces soft SUSY breaking terms 
 through the AMSB, and the resultant SUSY breaking mass scale is 
 characterized by $m_{\rm AMSB} \sim F_\phi/(16 \pi^2)$. 
Considering the anomaly mediation contribution to the soft scalar masses and A-terms, 
we take the minimal K\"ahler potential for the MSSM superfields, 
$K_{\rm MSSM} = Q_i^\dagger e^{2g_a V_a} Q_i$, where $Q_i$ stands for the MSSM
matter and Higgs superfields. 
Expanding $e^{K/3}$, the K\"ahler potential is described as 
\be 
{\cal L}_{\rm kin} = \int d^4 \theta \,\phi^\dagger \phi \,
 Q_i^\dagger e^{2g_a V_a} Q_i  + \cdots \; .
\label{matterK}
\ee
As discussed in Ref. \cite{method}, in softly broken supersymmetry, the soft terms associated to a chiral
superfield $Q_i$ can be collected in a running superfield wave function ${\cal Z}_i(\mu_R)$ such that
\be
\ln {\cal Z}_i(\mu_R) = \ln Z_i(\mu_R) + [A_i(\mu_R) \theta^2 + h.c.] - \widetilde{m}_i^2 (\mu_R) \theta^4 \;.
\ee
The running wave functions can be defined as $Z_i(\mu_R) = c_i(p^2 = -\mu_R^2)$, 
where $c_i$ is the coefficient of $Q_i^\dag Q_i$ in the one point-irreducible (1PI) effective action.
Therefore, turning on superconformal anomaly amounts to the shift $\mu_R \to \mu_R/(\phi^\dag \phi)^{1/2}$.
\be
{\cal Z}_i(\mu_R) = Z_i\left(\frac{\mu_R}{(\phi^\dag \phi)^{1/2}}\right)\;.
\ee
According to the method developed in Ref.~\cite{method} (see also Ref.~\cite{PR}), 
soft SUSY breaking terms (each gaugino masses $M_a$, sfermion squared masses
$\widetilde{m}_i^2$ and $A$-parameters) at the scale $\mu_R$ can 
be extracted from renormalized gauge kinetic functions and SUSY wave function renormalization coefficients, 
\bea 
&& M_{a}(\mu_R) = \frac{1}{16\pi^2} b_a g_a^2(\mu_R) F_{\phi} \; , 
 \nonumber\\ 
&& \widetilde{m}_i^2(\mu_R) = \frac{1}{2} \frac{d\gamma_i (\mu_R)}{d \ln \mu_R}  |F_{\phi}|^2 \; , 
 \nonumber \\
&& A_{ijk}(\mu_R) =- \left[ 
   \gamma_i (\mu_R) +\gamma_j (\mu_R) + \gamma_k (\mu_R) \right] F_{\phi} \; .
\label{softterms}
\eea
Here, 
$g_a$ are the gauge couplings, 
$b_a$ are beta function coefficients, and 
$\gamma_i \equiv -(1/2)d \ln Z / d \ln \mu$ are anomalous dimensions of the matter and Higgs superfields. 
%The results in Eq.~(\ref{softterms}) are true at any energy scale,
%and 
All the soft mass parameters can be described by only one parameter, 
$F_\phi$, so the anomaly mediation is highly predictive.

There are remaining two parameters in the Higgs sector, 
 namely $\mu$ and $B \mu$ terms, 
 that are responsible for electroweak symmetry breaking
 and should be of the order of the electroweak scale. 
%As in the anomaly mediation scenario, 
% the natural value of the $B$-parameter would be 
% $B \sim F_{\phi} \gg m_{\rm soft}$, and 
% the Higgs sector should be extended in order to 
% achieve the $B$-parameter being at the electroweak scale. 
%
Although some fine-tuning among parameters is necessary 
to realize $\mu \sim B \sim M_Z$,
in the following analysis we treat them as free parameters 
%That is, $\mu$ and $B \mu$ are replaced 
% into two free parameters $\tan \beta$ and ${\rm sgn}(\mu)$, 
so that the value of $|\mu|$ and $B \mu$ are determined by the stationary condition 
 of the Higgs potential. 

Let us consider the following superpotential.
\bea
W &=& -(Y_{U})_{ij} H_2 Q_i U^c_j  + (Y_{D})_{ij} H_1 Q_i D^c_j 
      -(Y_{\nu})_{ij} H_2 L_i N^c_j  + (Y_{E})_{ij} H_1 L_i E^c_j \nonumber\\
   && -\mu H_1 H_2 -\mu' \Delta_1 \Delta_2 +\frac{1}{2} f_{ij} \Delta_1 N^c_i N^c_j \;,
\eea
where $\Delta_1$ and $\Delta_2$ have $B-L$ charge $-2$ and $+2$ respectively.
Neglecting Yukawa couplings for first two generations, anomalous dimensions are given by
\bea
16\pi^2 \gamma_{Q_i} &=& -\frac{8}{3} g_3^2 - \frac{3}{2} g_2^2 - \frac{1}{18} g_Y^2 - \frac{2}{9} g_{B-L}^2
+( y_t^2+y_b^2) \delta_{i 3} \;, \nonumber \\
16\pi^2 \gamma_{U^c_i}&=&  -\frac{8}{3} g_3^2 - \frac{8}{9} g_Y^2 - \frac{2}{9} g_{B-L}^2
+ 2 y_t^2  \delta_{i 3} \; ,  \nonumber \\
16\pi^2 \gamma_{D^c_i} &=& -\frac{8}{3} g_3^2 - \frac{2}{9} g_Y^2 - \frac{2}{9} g_{B-L}^2
+ 2 y_b^2  \delta_{i 3} \; ,  \nonumber \\
16\pi^2 \gamma_{L_i} &=& -\frac{3}{2} g_2^2 - \frac{1}{2} g_Y^2 - 2 g_{B-L}^2
+ (y_{\nu}^2 + y_\tau^2)  \delta_{i 3} \; ,  \nonumber \\
16\pi^2 \gamma_{N^c_i} &=& - 2 g_{B-L}^2 + f^2 + 2 y_\nu^2 \delta_{i 3} \; ,  \nonumber \\
16\pi^2 \gamma_{E^c_i} &=& - 2 g_Y^2 - 2 g_{B-L}^2
+ 2 y_\tau^2 \delta_{i 3} \; ,  \nonumber \\
16\pi^2 \gamma_{H_1} &=& -\frac{3}{2} g_2^2 - \frac{1}{2} g_Y^2
+ 3 y_b^2 + y_\tau^2  \; , \nonumber \\
16\pi^2 \gamma_{H_2} &=& -\frac{3}{2} g_2^2 - \frac{1}{2} g_Y^2
+ 3 y_t^2 + y_{\nu}^2 \; , \nonumber \\
16\pi^2 \gamma_{\Delta_1} &=& -8 g_{B-L}^2 + f^2 \; , \nonumber \\
16\pi^2 \gamma_{\Delta_2} &=& -8 g_{B-L}^2 \; .
\eea

%Neglecting Yukawa couplings for first two generations, 
The soft scalar masses are explicitly written as
%Here $g_Y$ is the $U(1)_Y$ gauge coupling constant and is related to the GUT 
%normalized one as $g_1^2 = (5/3) g_Y^2$, 
\bea
m_{\widetilde{q}_i}^2 &=& \MSUSY^2
\left[
8 g_3^4 - \frac{3}{2} g_2^4 - \frac{11}{18} g_Y^4 -\frac{16}{3} g_{B-L}^4 
+ (y_t^2 b_{y_t} + y_b^2 b_{y_b}) \delta_{i3} 
\right] \;, \nonumber\\
m_{\widetilde{u}_i}^2 &=& \MSUSY^2
\left[
8 g_3^4 - \frac{88}{9} g_Y^4 -\frac{16}{3} g_{B-L}^4 
+  2 y_t^2 b_{y_t}  \delta_{i3}
\right] \;, \nonumber\\
m_{\widetilde{d}_i}^2 &=& \MSUSY^2
\left[
8 g_3^4 - \frac{22}{9} g_Y^4 -\frac{16}{3} g_{B-L}^4 
+ 2 y_b^2 b_{y_b} \delta_{i3}
\right]\;, \nonumber\\
m_{\widetilde{\ell}_{i}}^2 &=& \MSUSY^2
\left[
-\frac{3}{2} g_2^4 - \frac{11}{2} g_Y^4 -48 g_{B-L}^4 
+ (y_{\nu}^2 b_{y_\nu} + y_\tau^2 b_{y_\tau}) \delta_{i3} 
\right] \;, \nonumber\\
m_{\widetilde{\nu}_{i}}^2 &=&\MSUSY^2 
\left[ -48 g_{B-L}^4 + f^2 b_f + 2 y_{\nu}^2 b_{y_{\nu}} \delta_{i3} \right] \;, \nonumber\\
m_{\widetilde{e}_i}^2 &=&\MSUSY^2
\left[
- 22 g_Y^4 -48 g_{B-L}^4 
+ 2 y_\tau^2 b_{y_\tau} \delta_{i3}
\right] \;, \nonumber\\
m_{\widetilde{\Delta_1}}^2 &=&\MSUSY^2
\left[
-192 g_{B-L}^4 + f^2 b_f
\right] \;, \nonumber\\
m_{\widetilde{\Delta_2}}^2 &=&\MSUSY^2
\left[
-192 g_{B-L}^4
\right] \;.
\eea
%%%%%%%%%%%%%%%%%%%%%%%%%%%%%%%%%%%%%%%%%%
where $b_{y_t}$, $b_{y_b}$, $b_{y_{\nu}}$, $b_{y_\tau}$ and $b_f$ are given by 
\bea
b_{y_t} &=& 6 y_t^2 + y_b^2 + y_{\nu}^2
- \frac{16}{3} g_3^2 - 3g_2^2 - \frac{13}{9} g_Y^2 - \frac{4}{9} g_{B-L}^2 \;,
\nonumber\\
b_{y_b} &=& y_t^2 + 6 y_b^2 + y_\tau^2 
- \frac{16}{3} g_3^2 - 3g_2^2 - \frac{7}{9} g_Y^2 - \frac{4}{9} g_{B-L}^2 \;,
\nonumber\\
b_{y_\nu} &=& 3 y_t^2 + 4 y_\nu^2 + y_{\tau}^2 + f^2 - 3g_2^2 - g_Y^2 - 4 g_{B-L}^2 \;,
\nonumber\\
b_{y_\tau} &=& 3 y_b^2 + 4 y_\tau^2 + y_{\nu}^2 - 3g_2^2 - 3 g_Y^2 - 4 g_{B-L}^2 \;,
\nonumber\\
b_f &=& 4 y_{\nu}^2 + 3 f^2 - 12 g_{B-L}^2 \;.
\eea
Also, the Higgs soft masses are given by
\bea
m_{H_1}^2 &=& \MSUSY^2
\left[-\frac{3}{2} g_2^4 - \frac{11}{2} g_Y^4 + 3 y_b^2 b_{y_b} + y_\tau^2 b_{y_\tau} 
 \right] \;,
\nonumber\\
m_{H_2}^2 &=& \MSUSY^2
\left[-\frac{3}{2} g_2^4 - \frac{11}{2} g_Y^4 + 3 y_t^2 b_{y_t} 
\right] \;.
\eea
The Higgs mass parameters, $\mu$-term and $B \mu$-term, 
 are determined by the electroweak symmetry breaking conditions, 
\bea
|\mu|^2 &=& 
\frac{m_{H_1}^2 - m_{H_2}^2 \tan^2 \beta}{\tan^2 \beta -1}
- \frac{1}{2} M_Z^2  \;,
 \nonumber\\
B \mu &=& \frac{1}{2} 
 \left[m_{H_1}^2 + m_{H_2}^2  + 2 |\mu|^2 \right] \sin 2 \beta \; .
\label{mu}
\eea
The $A$-parameters in the AMSB scenario are given by
\bea
 A_{ijk} = - \left( \gamma_i +\gamma_j + \gamma_k \right) F_{\phi} \; 
\eea 
 with the above anomalous dimensions. 
Finally, the gaugino masses are given by 
\bea
M_{B-L} &=&  24 g_{B-L}^2 \MSUSY \; , \nonumber\\
M_1     &=&  11 g_Y^2 \MSUSY \; , \nonumber\\
M_2     &=& g_2^2 \MSUSY \; , \nonumber\\
M_3     &=& - 3 g_3^2 \MSUSY \; .
\eea
The mass ratios are approximately $M_{B-L} : M_1 : M_2 : M_3 = 57 g_{B-L}^2 : 3 : 1 : 10$.
So the Wino (rather than the more conventional Bino) is the lightest SUSY particle (LSP), 
and the gluino is an order of magnitude heavier than the LSP.
Those predictions for the gaugino masses in the AMSB, that is,
a Wino-like LSP, has interesting phenomenological consequences. 
The remarkable fact is that the lightest chargino mass is nearly degenerated 
with the lightest neutralino mass.

%%%%%%%%%%%%%%%%%%%%%%%%%%%%%%%%%%%%%%%%%%%%%%%%%%%%%%%
%%%%%%%%%%%%%%%%%%%%%%%%%%%%%%%%%%%%%%%%%%%%%%%%%%%%%%%
\section{Contributions from the Z-prime mediation}
%%%%%%%%%%%%%%%%%%%%%%%%%%%%%%%%%%%%%%%%%%%%%%%%%%%%%%%
%%%%%%%%%%%%%%%%%%%%%%%%%%%%%%%%%%%%%%%%%%%%%%%%%%%%%%%
Here we give a brief review of the Z-prime mediation of SUSY breaking \cite{ZMSB1, ZMSB2}
by discussing the pattern of the soft SUSY breaking parameters, 
the masses of the $\zp$-ino and of the MSSM squarks and gauginos,
which are the most robust predictions of this scenario.
At the SUSY breaking scale, $\Lambda_S$,
SUSY breaking in the hidden sector is assumed to generate a SUSY
breaking mass for the fermionic component of the ${\rm U}(1)_{\rm B-L}$ vector superfield.
Given details of the hidden sector, its value could be evaluated via the standard technique
of analytical continuation into superspace \cite{ArkaniHamed:1998kj}.
In particular, the gauge kinetic function of the field strength superfield ${\cal W}_{B-L}^\alpha$
at the SUSY breaking scale is  
\begin{eqnarray}
\label{eqn:zpinomass}
\mathcal{L}_{\tilde{Z}_{B-L}} &=& 
\int d^2 \theta 
\left[ 
\frac{1}{g_{B-L}^2} + \beta_{B-L}^{hid} \ln \left( \frac{\Lambda_S}{M} \right) \right. 
+\left. \beta^{vis}_{B-L}\ln \left( \frac{\Lambda_S}{M_{\tilde{Z}_{B-L}}} \right) 
\right] 
{\cal W}_{B-L}^\alpha {\cal W}_{B-L}^\alpha \;,
\end{eqnarray}
where $M$ is the messenger scale, which we have assumed to be around
the SUSY breaking scale, $M \sim
\Lambda_S$. $\beta_{B-L}^{hid}$ and $\beta^{vis}_{B-L}$ are
$\beta$-functions induced by ${\rm U}(1)_{\rm B-L}$ couplings to hidden and visible
sector fields, respectively.    Using analytical
continuation, we replace $M$
with $M+ \theta^2 F$, where $F$ is the SUSY breaking order
parameter.  We obtain the $\tilde{Z}_{B-L}$ mass as $ M_{\tilde{Z}_{B-L}} \sim g_{B-L}^2
\beta_{B-L}^{hid} F/M$.
We assume that the ${\rm U}(1)_{\rm B-L}$ gauge symmetry is
not broken in the hidden sector.
And we assume some sequestering mechanism so that only the $B-L$ gaugino 
obtains a leading order mass term while the threshold corrections to
the squrks and sleptons are only arisen at the next leading order as similar
to the case of the gaugino mediation, where the $B-L$ gaugino lives in the bulk
in a five dimesional setup while squarks and sleptons are put on the brane.
In such a case, only the $B-L$ gaugino obtains a mass while the scalar masses
receive negligible threshold corrections at the lowest order since they
receive volume suppression.

Since all the chiral superfields in the visible sector are
charged under ${\rm U}(1)_{\rm B-L}$, so all the corresponding scalars receive  soft
mass terms at 1-loop of order
\bea
\label{eqn:scalarmass}
m^2_{\tilde{q}_i} &=& \frac{8}{9} \frac{\alpha_{B-L}}{4 \pi} M_{\tilde{Z}_{B-L}}^2
\ln\left(\frac{\Lambda_S}{M_{\tilde{Z}_{B-L}}} \right),
\nonumber\\
m^2_{\tilde{\ell}_i} &=& 8\, \frac{\alpha_{B-L}}{4 \pi} M_{\tilde{Z}_{B-L}}^2
\ln\left(\frac{\Lambda_S}{M_{\tilde{Z}_{B-L}}} \right),
\eea
where $\alpha_{B-L}=g_{B-L}^2/(4\pi)$ and $ Q^f_{B-L}$ is
the ${\rm U}(1)_{\rm B-L}$ charge of $f$.

The MSSM gaugino masses, however,
can only be generated at 2-loop level since they do not directly couple to the ${\rm U}(1)_{\rm B-L}$,
\bea
\label{eqn:gauginomass}
M_a
&=& 4 c_a\, \frac{\alpha_{B-L}}{4 \pi} \frac{\alpha_a}{4 \pi} M_{\tilde{Z}_{B-L}}
\ln\left(\frac{\Lambda_S}{M_{\tilde{Z}_{B-L}}} \right) \;,
\eea
where $(c_1, c_2, c_3) = (\frac{92}{15}, 4,  \frac{4}{3})$.

From the discussion above, we see that the gauginos are considerably lighter
than the sfermions. 
Taking  ${m}_{\tilde{f}} \simeq 100$ - $1000$ GeV, we find
\begin{equation}
M_{\tilde{Z}_{B-L}} \simeq 10^4 ~\GeV
\end{equation}
 and then the $\zp$ mediated contribution is well-suppressed:
\begin{equation}
M_a \simeq  10^{-4} M_{\tilde{Z}_{B-L}} \simeq  1~\GeV\;,
\end{equation}
which can be negligible compared to the contributions from anomaly mediation.

%%%%%%%%%%%%%%%%%%%%%%%%%%%%%%%%%%%%%%%%%%%%%%%%%%%%%%%
%%%%%%%%%%%%%%%%%%%%%%%%%%%%%%%%%%%%%%%%%%%%%%%%%%%%%%%
\section{RGEs and its numerical evaluations}
%%%%%%%%%%%%%%%%%%%%%%%%%%%%%%%%%%%%%%%%%%%%%%%%%%%%%%%
%%%%%%%%%%%%%%%%%%%%%%%%%%%%%%%%%%%%%%%%%%%%%%%%%%%%%%%
Now we consider the RGEs and analyze
the running of the scalar masses $m_{\Delta_1}^2$ and
$m_{\Delta_2}^2$. The key point for implementing the radiative $B-L$
symmetry breaking is that the scalar potential $V(\Delta_1,\Delta_2)$
receives substantial radiative corrections \cite{Khalil:2007dr, KK}. 
In particular, a negative (mass)$^2$ would trigger the $B-L$ symmetry breaking. 
We argue that the masses of Higgs fields $\Delta_1$ and
$\Delta_2$ run differently in the way that $m^2_{\Delta_1}$ can be
negative whereas $m^2_{\Delta_2}$ remains positive. 
The RGE for the $B-L$ coupling and
mass parameters can be derived from the general results for SUSY
RGEs of Ref. \cite{Martin}.

For the RGEs of the Yukawa couplings, we consider to include the additional
contribution from the the $U(1)_{\rm B-L}$ gauge sector.
\bea
16 \pi^2 \frac{d\, y_A}{d \ln\mu} = b_{A} \, y_A \;,
\eea
where $A = (t,b,\nu,\tau,f)$, and $b_A$ is shown in the section~1.
The RGEs of the MSSM gauge couplings are the same as MSSM,
while the RGE of the $U(1)_{\rm B-L}$ gauge coupling is given by
\bea
16 \pi^2 \frac{d\, g_{B-L}}{d \ln\mu} = b_{B-L} \,g_{B-L}^3 \;,
\eea
where $b_{B-L} = 24$.
For the RGEs of the gaugino masses, it can be written as follows.
\bea
16 \pi^2 \frac{d\, M_{\tilde{Z}_{B-L}}}{d \ln\mu} &=& 
2 b_{B-L} g_{B-L}^3 M_{\tilde{Z}_{B-L}} \;, \nonumber\\
16 \pi^2 \frac{d\, M_a}{d \ln\mu} &=& 
\left[
\mbox{\sf MSSM + see-saw}
\right]
+ \frac{4 c_a g_a^2}{16 \pi^2} g_{B-L}^2 M_{\tilde{Z}_{B-L}} \;,
\eea
where $(c_a) = (92/15,4,4/3)$.
For the RGEs of the A-terms, it can be written as follows.
\bea
16 \pi^2 \mu \frac{d }{d \mu} \tilde{A}_A
&=&
\left[
\mbox{\sf MSSM + see-saw}
\right]
-2 a_A g_{\rm B-L}^2 (\tilde{A}_A - 2 M_{\tilde{Z}_{B-L}} Y_A ) \;,
\eea
where $\tilde{A}_A = A_A Y_A$ with $A = (t,b,\nu,\tau)$ and 
$(a_t, a_b, a_\nu, a_\tau) = (\frac{2}{9}, \frac{2}{9},2,2)$.
The RGE of the $A_f$-term can be written as
\bea
16 \pi^2 \mu \frac{d }{d \mu} \tilde{A}_f
&=&
\left(
9\, {\rm Tr} [f^\dag f]
+ 2 \, {\rm Tr} [Y_\nu^\dag Y_\nu] \right) \tilde{A}_f
+ 8\, f \,Y_\nu^\dag \tilde{A}_\nu \;.
\eea
The RGEs of the soft scalar masses are given by
\bea
16 \pi^2 \mu \frac{d m_{\Delta_1}^2}{d \mu}
&=&
2\, {\rm Tr} [f^\dag f] m_{\Delta_1}^2 + 4 \, {\rm Tr} [f^\dag m_N^2 f] 
- 32  g_{B-L}^2 |M_{\tilde{Z}_{B-L}}|^2 \;.
\nonumber\\
16 \pi^2 \mu \frac{d m_{\Delta_2}^2}{d \mu}
&=&
- 32  g_{B-L}^2 |M_{\tilde{Z}_{B-L}}|^2 \;.
\nonumber\\
16 \pi^2 \mu \frac{d m_{\tilde{f}}^2}{d \mu}
&=&
\left[
\mbox{\sf MSSM + see-saw}
\right]
- 8  g_{B-L}^2 (Q_{B-L}^{f})^2 |M_{\tilde{Z}_{B-L}}|^2 \;.
\eea
where $Q_{B-L}^{f}$ is the $B-L$ charge of each chiral multiplet 
$f = Q, U^c, D^c, L, N^c$.
For the RGEs of the $\mu'$-term, it can be written as follows.
\bea
16 \pi^2 \mu \frac{d }{d \mu} \mu'
= ({\rm Tr} [f^\dag f] - 16 g_{B-L}^2) \mu' \;.
\eea

In the numerical analysis, we fix $F_{\phi}$ to $10^5$ GeV for simplicity.
So we have only three free parameters,
\bea
g_{B-L}  \;,\;  f  \;,\;  M_{\tilde{Z}_B-L}  \;.
\eea
Once we fix $g_{B-L}$, $f$ and $M_{\tilde{Z}_B-L}$ at the SUSY breaking scale 
$\Lambda= \sqrt{ F_{\phi} M_{\mathrm{pl}} } \simeq 10^{11}$ GeV, 
all the soft SUSY breaking parameters due to AMSB and $Z^\prime$ mediation at $\Lambda$ 
are also fixed, and RGE evolutions provide us with informations at low scale.

\begin{figure}[htbp]
\begin{center}
\includegraphics[width=.8\linewidth]{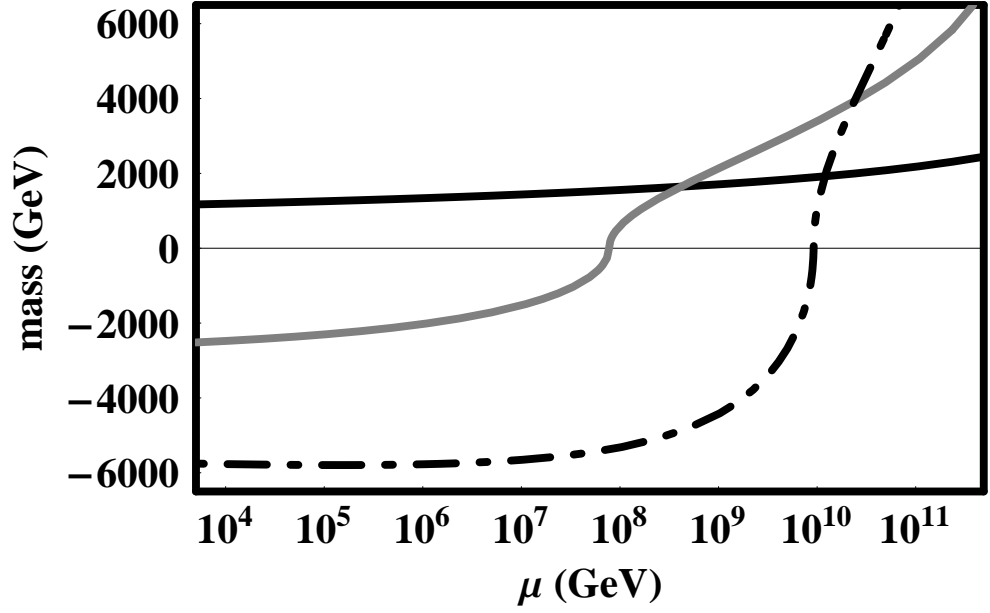}
\end{center}
\caption{
The evolution of the soft mass for the field $\Delta_1$
from the SUSY breaking scale to the $B-L$ gaugino mass scale.
The solid black, gray and dashed black lines are for 
$f=1.5,\ 2.5,\ 3.5$, respectively.
Here we have chosen $F_\phi=100$ TeV, $g_{B-L}=0.1$ and
$M_{\widetilde{Z}_{B-L}}=5$ TeV.
}
\label{fig:delta}
\end{figure}

Fig.~{\ref{fig:delta}} shows the evolutions of the soft mass for the field $\Delta_1$.
In Fig.~{\ref{fig:delta}}, from top to the bottom curves, 
we varied the value of $f$ as $f=1.5,\, 2.5,\, 3.5$ with 
%$F_\phi$, $g_{B-L}$ and $M_{\tilde{Z}_B-L}$ being fixed to 
$F_\phi=100$ TeV, $g_{B-L}=0.1$ and $M_{\widetilde{Z}_{B-L}}=5$ TeV.

For example, for the case of $f=2.5$, the soft mass squared for the fields $\Delta_1$
goes across the zeros toward negative value, that is nothing but the realization
of the radiative symmetry breaking of ${\rm U}(1)_{B-L}$ gauge symmetry.
The seesaw scale is found to be at $v_{B-L} = 10^4$ GeV. 
Hence the right-handed neutrinos obtain their masses of $M_N = f v_{B-L} \simeq \times 10^4$ GeV.
The running behavior in Fig.~{\ref{fig:delta}} can be understood in the following way.
Starting from the high energy scale, the soft mass squared increases
because of the gauge coupling contributions, and decrease of the mass squared
is caused by the Yukawa coupling $f$ that dominate over the gauge coupling contribution. 

\begin{figure}[htbp]
 \begin{minipage}{0.5\linewidth}
  \centering
  \includegraphics[width=7.9cm,clip]{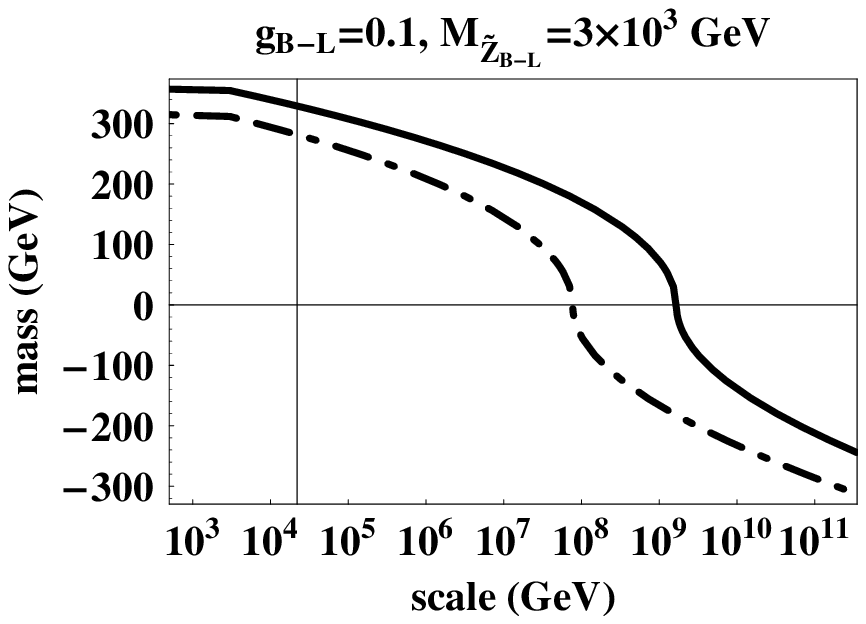}
 \end{minipage}
 \begin{minipage}{0.5\linewidth}
  \centering
  \includegraphics[width=7.9cm,clip]{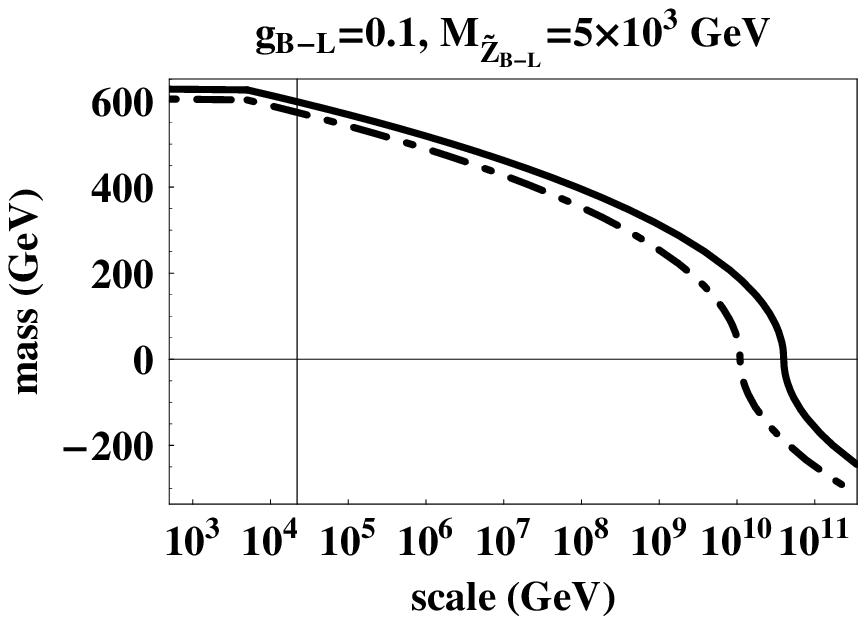}
 \end{minipage}
 \begin{minipage}{0.5\linewidth}
  \centering
  \includegraphics[width=7.9cm,clip]{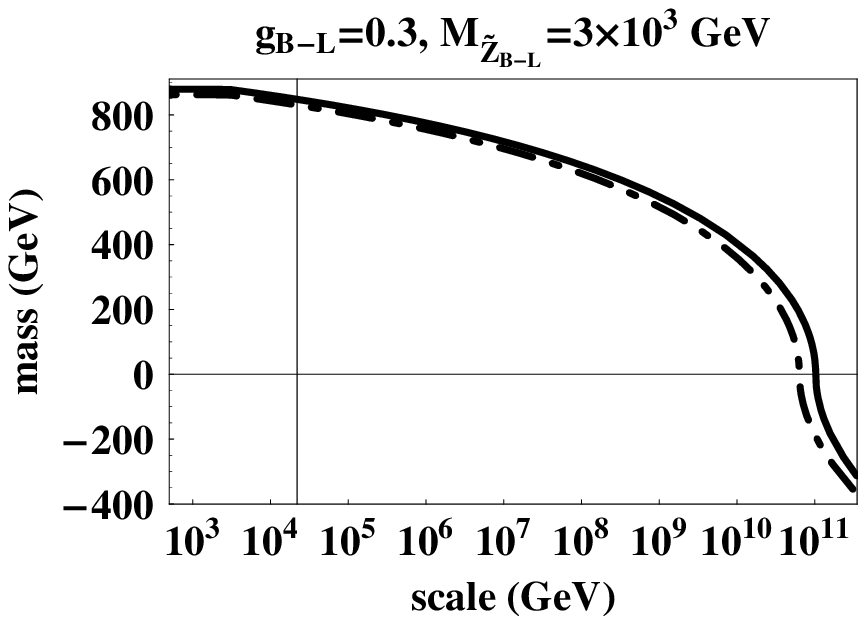}
 \end{minipage}
 \begin{minipage}{0.5\linewidth}
  \centering
  \includegraphics[width=7.9cm,clip]{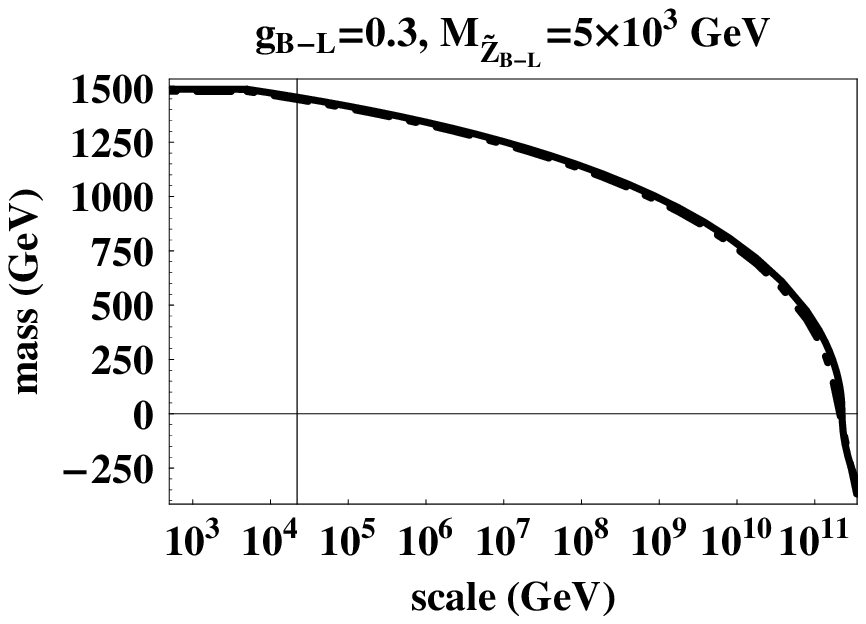}
 \end{minipage}
 \begin{minipage}{0.5\linewidth}
  \centering
  \includegraphics[width=7.9cm,clip]{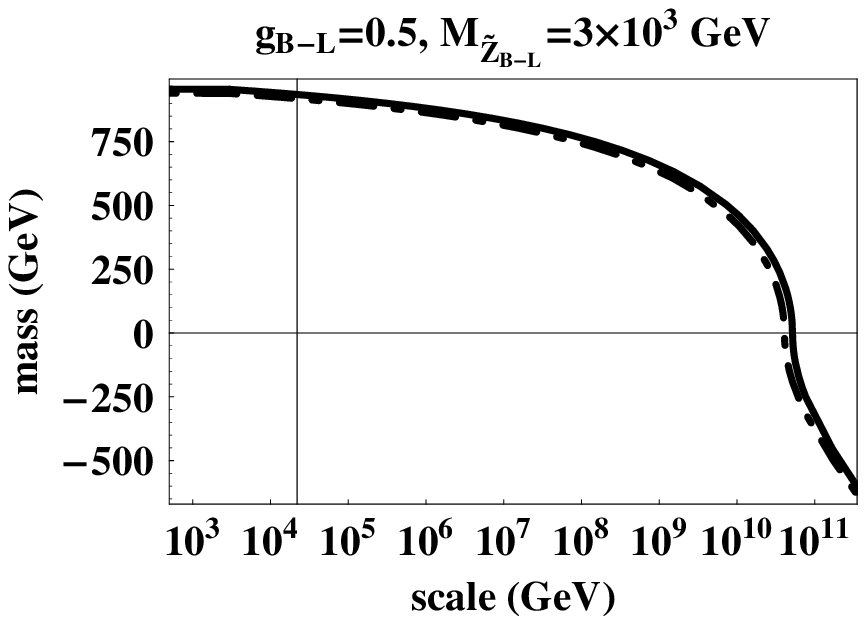}
 \end{minipage}
 \begin{minipage}{0.5\linewidth}
  \centering
  \includegraphics[width=7.9cm,clip]{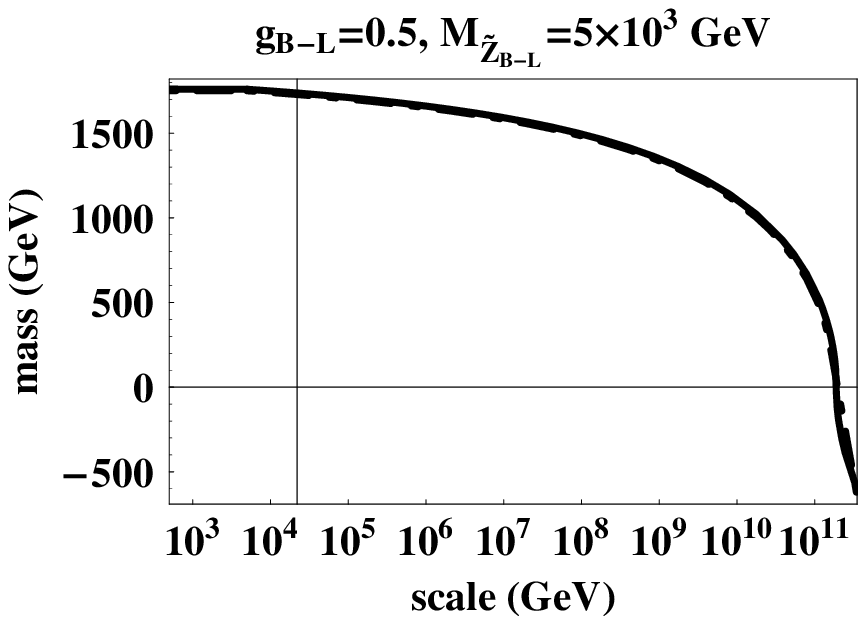}
 \end{minipage}
 \caption{\small The running behavior of the soft mass parameters 
 $m_{\tilde{\ell}_1}$ (solid line) and 
 $m_{\tilde{e}_1}$ (dotted line) are shown.
Here we have chosen $F_\phi=50$ TeV and $f=2.5$.
}    
 \label{fig:slepton} 
\end{figure}

Fig.~{\ref{fig:slepton}} show the evolutions of the soft mass for 
sleptons, where the Yukawa coupling $f$ is fixed to $2.5$, 
since their spectra are almost independent of the value of $f$.
As seen in Fig.~{\ref{fig:slepton}}, the larger $M_{\tilde{Z}_B-L}$ gives 
the more positive slepton mass. 
This behavior is easily understood from Eq.~(\ref{slepton}), the RGE of the slepton.
On the other hand, the larger $g_{B-L}$ gives the degenerate mass spectra.
This is because, 
$ m_{\widetilde{\ell}_{1,2}}^2 $ and $ m_{\widetilde{e}_{1,2}}^2 $ at $\Lambda$ 
depend only on $g_{B-L}$ in the case of the large $g_{B-L}$. 
These degenerate mass spectra are one of the outstanding feature of this scenario.

In Table~\ref{table}, we show some example data of the resultant sparticle mass spectrum and Higgs boson masses,
where we took $\tan \beta=10$, $F_\phi=50$ TeV and $f=2.5$.
Here, the standard model-like Higgs boson mass 
 is evaluated by including one-loop corrections 
 through top and scalar top quarks, 
\bea
 \Delta m_h^2 = \frac{3}{4 \pi^2} y_t^4 v^2 \sin^4 \beta
 \ln \left(\frac{m_{\tilde{t}_1} m_{\tilde{t}_2}}{m_t^2} \right) \;, 
\eea 
which is important to push up the Higgs boson mass 
 so as to satisfy the LEP II experimental bound, $m_h \gtrsim 114$ GeV. 
As can be understood from the RGEs and 
 the soft SUSY breaking parameters %at the GUT scale 
 presented in the previous section, 
 the resultant soft SUSY breaking parameters are 
 proportional to $F_\phi$.
Thus, as we take $F_\phi$ larger, 
 sparticles become heavier and, 
 accordingly, Higgs boson masses become larger.

%%%%%%%%%%%%%%%%%%%%%%%%%%%%%%%%%%%%%%%%%%%%%%%%%%
\section{Dark matter relic density}
%%%%%%%%%%%%%%%%%%%%%%%%%%%%%%%%%%%%%%%%%%%%%%%%%%
In this section we discuss the cosmological features of the lightest neutralino.
The recent Wilkinson Microwave Anisotropy Probe (WMAP) satellite data \cite{WMAP5}  provide
estimations of various cosmological parameters with greater accuracy. 
The current density of the universe is composed of about 73\% of dark energy and
27\% of matter. Most of the matter density is in the form of the CDM, 
and its density is estimated to be \cite{WMAP5}
\begin{eqnarray}
\Omega_{\rm CDM} h^2  = 0.1143 \pm 0.0034  \;. 
 \label{WMAP} 
\end{eqnarray}
If the R-parity is conserved in SUSY models, the LSP is stable. 
The lightest neutralino, if it is the LSP, is the plausible candidate for the CDM.

In the AMSB scenario or its extension with $Z^\prime$ mediation, the lightest neutralino is mostly Wino-like, 
and it undergoes rapid annihilation though reaction:
$\widetilde{W} \widetilde{W} \to W^+ W^-$. 
The resultant relic abundance is too small, which can roughly be estimated to be \cite{AMSB2}
\be
\Omega_{\widetilde{W}} h^2 \simeq 5 \times10^{-4} \, 
\left(\frac{M_{\widetilde{W}}}{100~\mbox{GeV}} \right)^2 \; . 
\label{WMAP2}
\ee
So the mass of the DM neutralino has to be very heavy to satisfy the WMAP data.
If the Wino-like neutralino with SU(2)$_L$ charge is much heavier than the weak gauge boson as described above, 
the weak interaction is a long-distance force for non-relativistic two-bodies states of such particles. 
If this non-perturbative effect (namely, Sommerfeld enhancement) of the dark matter at the freeze-out
temperature is taken into account, the abundance can be reduced by about 50\% \cite{non-pert1, Hisano:2006nn}.
Therefore, the allowed region exists for large value of $F_\phi$.

Such a large value of soft mass is disfavored in view of the little hierarchy problem.
In order to keep the neutralino DM light, non-thermal production of the DM should be considered
as proposed in \cite{Moroi-Randall}. 
Once we accept the non-thermal production of the LSP neutralino from the moduli decays,
then it is possible to produce sufficient relic abundance of the LSP neutralino
even for the light Wino-like neutralino DM.

%%%%%%%%%%%%%%%%%%%%%%%%%%%%%%%%%%%%%%%%
\section{Summary and discussion}
%%%%%%%%%%%%%%%%%%%%%%%%%%%%%%%%%%%%%%%%
Anomaly mediation of supersymmetry breaking (AMSB) is very attractive 
because the resultant soft supersymmetry breaking parameters 
at a given energy scale are determined only by physics at that energy scale
(UV insensitivity) and hence is highly predictive (only one parameter, $F_\phi$). 
However, there is the so-called tachyonic slepton problem.
In this paper, we have constructed a viable anomaly mediation scenario of SUSY breaking
by adding a contribution from the $Z^\prime$ mediated SUSY breaking contributions.
In the $Z^\prime$ mediated SUSY breaking scenario, while the scalar masses are generated
at the 1-loop level, however, gaugino masses can only be generated at 2-loop level, 
so the gaugino masses are completely determined by the pure anomaly mediation itself.
Therefore, the characteristic signature of the present model predictions appear in
the scalar partners mass spectra.
We have investigated the scalar partners mass spectra for several choices of parameters
in this model, for instance, for different values of the $Z^\prime$ gaugino mass.
The resultant sparticle mass spectra was found to be interesting in scope of the LHC.

\section*{Acknowledgments}
T.~Kikuchi would like to thank K.S. Babu for his hospitality
at Oklahoma State University.
The work of T.~Kikuchi is supported by the Research
Fellowship of the Japan Society for the Promotion of Science (\#1911329).
T.~Kubo would like to thank Y.~Okada for his encouragement.
We thank M.~Nagai and N.~Okada for their stimulating discussions.

\begin{appendix}
\section{RGEs in the MSSM with right-handed neutrinos}
%%%%%%%%%%%%%%%%%%%%%%%%%%%%%%%%%%%%%%%%%%%%%%%%%%%%%%%%%%%
\subsection{The 2-loop RGE for the gauge couplings}
%%%%%%%%%%%%%%%%%%%%%%%%%%%%%%%%%%%%%%%%%%%%%%%%%%%%%%%%%%%
\bea
16\pi^2 \mu \frac{d}{d \mu} ~g_1 &\!=\!& \frac{33}{5}\, g_1^3 
+ \frac{g_1^3}{16\pi^2} \left(\frac{199}{25} g_1^2 + \frac{27}{5} g^2_2
+ \frac{88}{5} g_3^2\right) \nn, \\
16\pi^2 \mu \frac{d}{d \mu} ~g_2 &\!=\!& \, g_2^3 + 
\frac{g_2^3}{16\pi^2} \left(\frac{9}{5} g_1^2 + 25 g^2_2
+ 24 g_3^2\right) \nn, \\
16\pi^2 \mu \frac{d}{d \mu} ~g_3 &\!=\!& -3 \, g_3^3 
+ \frac{g_3^3}{16\pi^2} \left(\frac{1}{5} g_1^2 + 9 g^2_2
+ 14 g_3^2\right) \nn.
\eea
Here $g_2 \equiv g$ is the $SU(2)_L$ gauge coupling constant and 
$g_1 \equiv \sqrt{\frac{5}{3}} g^\prime$ is the $U(1)$ gauge coupling 
constant with the GUT normalization 
($g_1 = g_2 = g_3$ at $\mu = M_{\rm GUT}$). 

%%%%%%%%%%%%%%%%%%%%%%%%%%%%%%%%%%%%%%%%%%%%%%%%%%%%%%%%%%
\subsection{The 1-loop RGE for the Yukawa couplings}
%%%%%%%%%%%%%%%%%%%%%%%%%%%%%%%%%%%%%%%%%%%%%%%%%%%%%%%%%%
\bea
16\pi^2 \mu \frac{d}{d \mu} Y_{u} &\!=\!& 
Y_u \left[
\left \{ -\frac{13}{15} g_1^2 - 3 g_2^2 -\frac{16}{3} g_3^2 
  + 3 \,{\rm Tr} ( Y_u^{\dagger} Y_u )
  +   {\rm Tr} ( Y_\nu^{\dagger} Y_\nu ) \right \}
{\bf 1}_{3\times3}
\right.
\nonumber \\
&& \left.\; + 3 \, ( Y_u^{\dagger} Y_u) 
      + (Y_{d}^{\dagger} Y_{d}) \right]\nn, \\
16\pi^2 \mu \frac{d}{d \mu} Y_{d} &\!=\!& 
Y_d \left[
\left \{ -\frac{7}{15} g_1^2 - 3 g_2^2 -\frac{16}{3} g_3^2 
  + 3 \,{\rm Tr} ( Y_d^{\dagger} Y_d )
  +   {\rm Tr} ( Y_e^{\dagger} Y_e ) 
\right \} {\bf 1}_{3\times3}
\right.
\nonumber \\
&& \left.
+ \; 3 \, ( Y_d^{\dagger} Y_d) 
+ (Y_{u}^{\dagger} Y_{u}) \right]\nn, \\
16\pi^2 \mu \frac{d}{d \mu} Y_{\nu} &\!=\!&
Y_\nu \left[
\left \{ -\frac{3}{5} g_1^2 - 3 g_2^2 
+ 3 \,{\rm Tr} \left( Y_{u}^{\dagger} Y_{u} \right)
+   {\rm Tr} \left( Y_{\nu}^{\dagger} Y_{\nu} \right) 
\right \}  {\bf 1}_{3\times3}
\right.
\nonumber \\
&& \left.
+ 3 \, \left( Y_{\nu}^{\dagger} Y_{\nu} \right)
+ \left(Y_{e}^{\dagger} Y_{e} \right) \right] 
\nn, \\
16\pi^2 \mu \frac{d}{d \mu} Y_{e} &\!=\!& 
Y_e \left[
\left \{ -\frac{9}{5} g_1^2 - 3 g_2^2 
  + 3 \,{\rm Tr} ( Y_d^{\dagger} Y_d)
  +   {\rm Tr} ( Y_e^{\dagger} Y_e) 
\right \} {\bf 1}_{3\times3}
\right.
\nonumber \\
&& \left.
+ 3 \, \left( Y_e^{\dagger} Y_e \right) + 
\left(Y_{\nu}^{\dagger} Y_{\nu} \right) \right]\nn. 
\eea
%

%%%%%%%%%%%%%%%%%%%%%%%%%%%%%%%%%%%%%%%%%%%%%%%%%%%%%%%%%%%%%
\subsection{The 2-loop RGE for the gaugino masses}
%%%%%%%%%%%%%%%%%%%%%%%%%%%%%%%%%%%%%%%%%%%%%%%%%%%%%%%%%%%%
\bea
16\pi^2 \mu \frac{d}{d \mu} M_1 &\!=\!& \frac{66}{5} \, g_1^2 M_1 
\nonumber\\
&\!+\!&
\frac{2 g_1^2}{16\pi^2} \left\{ \frac{199}{5} g^2_1 \left( 2 M_1  
\right)  + \frac{27}{5} g_2^2 \left( M_1 + M_2 \right) 
+ \frac{88}{5} g_3^2 \left( M_1 + M_3 \right)\right\} \nn, \\
16\pi^2 \mu \frac{d}{d \mu} M_2 &\!=\!& 2 \, g_2^2 M_2 
\nonumber\\
&\!+\!&
\frac{2 g_2^2}{16\pi^2} \left\{ \frac{9}{5} g^2_1 \left(M_1 + M_2 \right)
+ 25 g_2^2 \left( 2 M_2\right) + 24 g_3^2 
\left( M_2 + M_3 \right) \right\} \nn, \\
16\pi^2 \mu \frac{d}{d \mu} M_3 &\!=\!& -6\, g_3^2 M_3 
\nonumber\\
&\!+\!&
\frac{2 g_3^2}{16\pi^2} \left\{ \frac{11}{5} g^2_1 \left(M_1 + M_3 \right)
+ 9 g_2^2 \left( M_2 + M_3 \right) + 14 g_3^2 
\left( 2 M_3 \right) \right\} \nn.
\eea
%%%%%%%%%%%%%%%%%%%%%%%%%%%%%%%%%%%%%%%%%%%%%%%%%%%%%%%%%%%%
\subsection{The 1-loop RGE for the soft SUSY breaking mass terms}
%%%%%%%%%%%%%%%%%%%%%%%%%%%%%%%%%%%%%%%%%%%%%%%%%%%%%%%%%%%%%
\bea
16\pi^2 \mu \frac{d}{d \mu} \left( m^2_{\tilde{\sss q}} \right)_{ij} 
&\!=\!&
- \left( \frac{2}{15} g_1^2 \left| M_1 \right|^2 
+ 6 g_2^2 \left| M_2 \right|^2 + \frac{32}{3} g_3^2 \left| M_3 
\right|^2\right) \delta_{ij}
+ \frac{1}{5} g_1^2~S~\delta_{ij} \nonumber \\
&\!+\!& 
\left( m^2_{\tilde{\sss q}} Y_u^{\dagger} Y_u 
+ m^2_{\tilde{\sss q}} Y_d^{\dagger} Y_d 
+ Y_u^{\dagger} Y_u m^2_{\tilde{\sss q}} 
+ Y_d^{\dagger} Y_d  m^2_{\tilde{\sss q}} \right)_{ij} \nonumber \\
&\!+\!&
2 \left( Y_u^{\dagger} m^2_{\tilde u} Y_u
           + {m}^2_{H_u} Y_u^{\dagger} Y_u
+ A_u^{\dagger} A_u \right)_{ij} \nonumber \\
&\!+\!&
2 \left( Y_d^{\dagger} m^2_{\tilde {d}} Y_{d}
+ {m}^2_{H_d} Y_{d}^{\dagger} Y_{d}
+ A_{d}^{\dagger} A_{d} \right)_{ij}\nn, \\
%%%%%%%%%%%%%%%%%%%%%%%%%%%%%%%%%%%%%%%%%%%%%%%%%%%%%%%
16\pi^2 \mu \frac{d}{d \mu} \left( m^2_{\tilde{u}} \right)_{ij} 
&\!=\!&
- \left( \frac{32}{15} g_1^2 \left| M_1 \right|^2 
+ \frac{32}{3} g_3^2 \left| M_3 \right|^2\right) \delta_{ij}
- \frac{4}{5} g_1^2~S~\delta_{ij} \nonumber \\
&\!+\!&
 2 \left( m^2_{\tilde u} 
Y_u^{\dagger} Y_u + Y_u^{\dagger} Y_u m^2_{\tilde u} \right)_{ij} 
\nonumber \\
&\!+\!&
 4 \left( Y_u m^2_{\tilde{\sss q}} Y_u^{\dagger} + {m}^2_{H_u} 
Y_u^{\dagger} Y_u + A_u A_u^{\dagger} \right)_{ij}\nn, \\
16\pi^2 \mu \frac{d}{d \mu} \left( m^2_{\tilde{d}} \right)_{ij} 
&\!=\!&
- \left( \frac{8}{15} g_1^2 \left| M_1 \right|^2 
+ \frac{32}{3} g_3^2 \left| M_3 \right|^2\right) \delta_{ij}
+ \frac{2}{5} g_1^2~S~\delta_{ij} \nonumber \\
&\!+\!& 
2 \left( m^2_{\tilde d} 
Y_d^{\dagger} Y_d + Y_d^{\dagger} Y_d m^2_{\tilde d} \right)_{ij} 
\nonumber \\
&\!+\!&
4 \left( Y_d m^2_{\tilde{\sss q}} Y_d^{\dagger} + {m}^2_{H_d} 
Y_d^{\dagger} Y_d + A_d A_d^{\dagger} \right)_{ij}\nn, \\
%
%%%%%%%%%%%%%%%%%%%%%%%%%%%%%%%%%%%%%%%%%%%%%%%%%%%%%%%
%
16\pi^2 \mu \frac{d}{d \mu} \left( m^2_{\tilde{\sss \ell}} 
\right)_{ij}&\!=\!&   -\left( \frac{6}{5} g_1^2 \left| M_1 \right|^2 
+ 6 g_2^2 \left| M_2 \right|^2 \right) \delta_{ij}
-\frac{3}{5} g_1^2~S~\delta_{ij} \nonumber \\
&\!+\!&
\left( m^2_{\tilde{\sss \ell}} Y_e^{\dagger} Y_e 
+ m^2_{\tilde{\sss \ell}} Y_\nu^{\dagger} Y_\nu 
+ Y_e^{\dagger} Y_e m^2_{\tilde{\sss \ell}} 
+ Y_\nu^{\dagger} Y_\nu  m^2_{\tilde{\sss \ell}} \right)_{ij} \nonumber \\
&\!+\!&
2 \left( Y_e^{\dagger} m^2_{\tilde e} Y_e
           +{m}^2_{H_d} Y_e^{\dagger} Y_e
+ A_e^{\dagger} A_e \right)_{ij} \nonumber \\
&\!+\!&
2 \left( Y_{\nu}^{\dagger} m^2_{\tilde {\sss\nu}} Y_{\nu}
+ {m}^2_{H_u} Y_{\nu}^{\dagger} Y_{\nu}
+ A_{\nu}^{\dagger} A_{\nu} \right)_{ij}\nn, 
\label{slepton}
\\
16\pi^2 \mu \frac{d}{d \mu} \left( m^2_{\tilde{e}} \right)_{ij}&\!=\!& 
- \frac{24}{5} g_1^2 \left| M_1 \right|^2 \delta_{ij}
+ \frac{6}{5} g_1^2~S~\delta_{ij} + 2 \left( m^2_{\tilde e} 
Y_e^{\dagger} Y_e + Y_e^{\dagger} Y_e m^2_{\tilde e} \right)_{ij} 
\nonumber \\
&\!+\!&
4 \left( Y_e m^2_{\sss\widetilde \ell} Y_e^{\dagger} + {m}^2_{H_d} 
Y_e^{\dagger} Y_e + A_e A_e^{\dagger}\right)_{ij}\nn, \\
16\pi^2 \mu \frac{d}{d \mu} \left( m^2_{\tilde{\nu}} \right)_{ij} &\!=\!& 
2 \left( m^2_{\tilde \nu} Y_{\nu}^{\dagger} Y_{\nu} 
+ Y_{\nu}^{\dagger} Y_{\nu} m^2_{\tilde \nu} \right)_{ij} 
+ 4 \left( Y_{\nu} m^2_{\tilde \ell} Y_{\nu}^{\dagger}
+ {m}^2_{H_u} Y_{\nu}^{\dagger} Y_{\nu}
+ A_{\nu} A_{\nu}^{\dagger}\right)_{ij} \nn. 
\eea
\bea
{16 \pi^2} \mu \frac{d}{d \mu} (m^2_{H_u}) &=&  -\left(\frac{6}{5} 
g^{2}_{1} \abs{M_{1}}^{2} + 6 g^{2}_{2} \abs{M_{2}}^{2}\right)
+ \frac{3}{5} g_1^2 S
\nonumber\\
&\!+\!& 
6\, {\rm {\rm Tr}}\left(
m^2_{\tilde q} Y^{\dagger}_{u} Y_{u} + Y^{\dagger}_{u}
( m^2_{\tilde u} + m^2_{H_u} ) Y_{u} + A^{\dagger}_{u} A_{u}
\right) \nonumber\\
&\!+\!& 2\, {\rm Tr} \left( m^2_{\tilde{\sss \ell}} Y^{\dagger}_{\nu} Y_{\nu}
+ Y^{\dagger}_{\nu} ( m^2_{\tilde \nu} + m^2_{H_u}) Y_{\nu} 
+ A^{\dagger}_{\nu} A_{\nu} \right) \nn,\\
%Hd=h_d
{16 \pi^2} \mu \frac{d}{d \mu} (m^2_{H_d}) &=& -\left( \frac{6}{5} 
g^{2}_{1} \abs{M_{1}}^{2} + 6 g^{2}_{2} \abs{M_{2}}^{2}\right) 
- \frac{3}{5} g_1^2 S \nonumber\\
&\!+\!&
6\, {\rm Tr}\left( m^2_{\tilde q} Y^{\dagger}_{d} Y_{d} 
+ Y^{\dagger}_{d} 
( m^2_{\tilde d} + m^2_{H_d}) Y_{d} + A^{\dagger}_{d} A_{d}\right)
\nonumber\\
&\!+\!& 2\, {\rm Tr} \left( m^2_{\tilde{\sss \ell}} Y^{\dagger}_{e} Y_{e} 
+ Y^{\dagger}_{e} ( m^2_{\tilde e} + m^2_{H_d}) Y_{e} 
+ A_{e}^{\dagger} A_{e} \right) \nn,
\eea
where
\begin{equation}
S \equiv {\rm Tr} (m_{\tilde q}^2 + m_{\tilde d}^2 - 2 m_{\tilde u}^2
- m_{\tilde \ell}^2 + m_{\tilde e}^2 ) - {m}^2_{H_d} 
+ {m}^2_{H_u} \nn.
\end{equation}
%%%%%%%%%%%%%%%%%%%%%%%%%%%%%%%%%%%%%%%%%%%%%%%%%%%%%%%%%%%%
\subsection{The 1-loop RGE for the soft SUSY breaking A-terms}
%%%%%%%%%%%%%%%%%%%%%%%%%%%%%%%%%%%%%%%%%%%%%%%%%%%%%%%%%%%%%
\bea
16\pi^2 \mu \frac{d}{d \mu} A_{u_{ij}}&=&
\left\{ -\frac{13}{15} g_1^2 - 3 g_2^2 - \frac{16}{3} g_3^2 + 3 {\rm Tr} 
(Y_u^{\dagger} Y_u) + {\rm Tr}( Y_{\nu}^{\dagger} Y_{\nu}) 
\right \} A_{u_{ij}} 
\nonumber \\
&\!+\!&
2 \left\{ \frac{13}{15} g_1^2 M_1 + 3 g_2^2 M_2 +\frac{16}{3} g_3^2 M_3
+ 3 {\rm Tr}(Y_u^{\dagger} A_u)
+ {\rm Tr}( Y_{\nu}^{\dagger} A_{\nu}) \right \} Y_{u_{ij}}
\nonumber \\
&\!+\!& 
4 ( Y_u^{\dagger} Y_u A_u)_{ij} 
+ 5 ( A_u Y_u^{\dagger} Y_u)_{ij} 
+ 2 ( Y_u Y_d^{\dagger} A_d)_{ij} 
+ ( A_u Y_d^{\dagger} Y_d)_{ij} \nn,\\
%Ad
16 \pi^2 \mu \frac{d}{d \mu} A_{d_{ij}}&=&
\left\{
-\frac{7}{15} g_1^2 - 3 g_2^2  -\frac{16}{3} g_3^2 
+ 3 {\rm Tr} ( Y_d^{\dagger} Y_d)
+ {\rm Tr}( Y_{e}^{\dagger} Y_{e}) \right \} A_{d_{ij}}
\nonumber \\
&\!+\!& 2 \left\{ \frac{7}{15} g_1^2 M_1 + 3 g_2^2 M_2 + 
\frac{16}{3} g_3^2 M_3 + 3 {\rm Tr}( Y_d^{\dagger} A_d) 
+ {\rm Tr} ( Y_{e}^{\dagger} A_{e}) \right \} Y_{d_{ij}}
\nonumber \\ 
&\!+\!&
4 ( Y_d^{\dagger} Y_d A_d)_{ij} 
+ 5 ( A_d Y_d^{\dagger} Y_d)_{ij} 
+ 2 ( Y_d Y_u^{\dagger} A_u)_{ij} 
+ ( A_d Y_u^{\dagger} Y_u)_{ij} \nn, 
\\
16\pi^2 \mu \frac{d}{d \mu} A_{e_{ij}}&\!=\!&  
 \left\{ -\frac{9}{5} g_1^2 -3 g_2^2
+ 3 {\rm Tr} ( Y_d^{\dagger} Y_d )
+   {\rm Tr} ( Y_e^{\dagger} Y_e ) \right \} A_{e_{ij}} \nonumber \\
&\!+\!&
2 \left\{
\frac{9}{5} g_1^2 M_1 + 3 g_2^2 M_2 
+ 3 {\rm Tr} ( Y_d^{\dagger} A_d)
+   {\rm Tr} ( Y_e^{\dagger} A_e) \right \} Y_{e_{ij}} \nonumber \\
&\!+\!&
4 \left( Y_e^{\dagger} Y_e A_e \right)_{ij} 
+ 5 \left( A_e Y_e^{\dagger} Y_e \right)_{ij} 
+ 2 \left( Y_e Y_{\nu}^{\dagger} A_{\nu} \right)_{ij} 
+  \left( A_e Y_{\nu}^{\dagger} Y_{\nu} \right)_{ij} \nn, 
\\
16\pi^2 \mu \frac{d}{d \mu} A_{\nu_{ij}}&\!=\!&  
\left\{ -\frac{3}{5} g_1^2 -3 g_2^2 
+ 3 {\rm Tr} ( Y_u^{\dagger} Y_u)
+  {\rm Tr} ( Y_{\nu}^{\dagger} Y_{\nu}) \right \} A_{\nu_{ij}} \nonumber \\
&\!+\!&
2 \left\{ \frac{3}{5} g_1^2 M_1 + 3 g_2^2 M_2 
+ 3 {\rm Tr} ( Y_u^{\dagger} A_u)
+   {\rm Tr} ( Y_{\nu}^{\dagger} A_{\nu}) \right \} Y_{\nu_{ij}} 
\nonumber \\
&\!+\!&
 4 ( Y_{\nu}^{\dagger} Y_{\nu} A_{\nu})_{ij}
+ 5 ( A_{\nu} Y_{\nu}^{\dagger} Y_{\nu})_{ij} 
+ 2 ( Y_{\nu} Y_e^{\dagger} A_e)_{ij}
+ ( A_{\nu} Y_e^{\dagger} Y_e)_{ij} \nn. 
\eea
\end{appendix}

%\newpage
%%%%%%%%%%%%%%%%%%%%%%%%%%%%%%%%%%%%%%

%%%%%%%%%%%%%%%%%%%%%%%%%%%%%%%%%%%%%%%%%%%%%

%%%%%%%%%%%%%%%%%%%%%%%%%%%%%%%%%%%%%%%%%%%%%%
\newpage
%%%%%%%%%%%%%%%%%%%%%%%%%%%%%%%%%%%%%%%%%%%%%%
%%%%%%%%%%%%%%%%%%%%%%%%%%%%%%%%%
%%%%%%%%%%%%%%%%%%%%%%%%%%%%%%%%%
%%%%%%%%%%%%%%%%%%%%%%%%%%%%%%%%
%%%%%%%%%%%%%%%%%%%%%%%%%%%%%%%%%%
\pagestyle{empty}
%%%%%%%%%%%%%%%%%%%%%%%%%%%%%%%%%%
%\begin{center}
\begin{table}[htbp]
\centering
\begin{tabular}{|c|c|c|c|}
\hline \hline
$(g_{B-L},~ M_{\widetilde{Z}_{B-L}})$  & ($0.1$,~ $3$ TeV)
                                & ($0.3$, ~ $3$ TeV)
                                & ($0.5$,~ $3$ TeV)     \\
\hline \hline
$m_{\tilde{\chi}^0_{1,2,3,4}}$ & 132, 455, 719, 726
                               & 131, 455, 742, 749
                               & 131, 454, 745, 754 \\
$m_{\tilde{\chi}^{\pm}_{1,2}}$ & 133, 717
                               & 132, 741
                               & 132, 746 \\
$m_{\tilde{g}}$          &  1297
                         &  1298
                         &  1299 \\
\hline
$m_{{\tilde{e},\tilde{\mu}}_{1,2}}$
                          & 318, 360 
                          & 864, 881  
                          & 941, 957  \\
$m_{\tilde{\tau}_{1,2}}$  & 299, 355 
                          & 855, 877  
                          & 931, 953  \\
\hline
$m_{{\tilde{u},\tilde{c}}_{1,2}}$ 
                          & 1216, 1228 
                          & 1246, 1257 
                          & 1252, 1263  \\
$m_{\tilde{t}_{1,2}}$     & 979, 1121 
                          & 1004, 1146 
                          & 1007, 1149 \\
\hline
$m_{{\tilde{d},\tilde{s}}_{1,2}}$ 
                          &1219, 1226 
                          &1248, 1256
                          &1255, 1262  \\
$m_{\tilde{b}_{1,2}}$     & 1088, 1211 
                          & 1115, 1240
                          & 1119, 1247   \\
\hline
$m_h$                    & 124
                         & 124
                         & 124 \\
$m_H$                    & 663
                         & 685
                         & 690 \\
$m_A$                    & 662
                         & 685
                         & 690 \\
$m_{H^{\pm}}$            & 667
                         & 690
                         & 694 \\
\hline \hline
\end{tabular}
%\caption{
%Sparticle and Higgs boson mass spectra (in units of GeV) 
% in the case of $\tan \beta=10$ and $F_\phi=50$ TeV.}
\label{table1}
\vspace{1cm}
%\end{table}
%\end{center}
%%%%%%%%%%%%%%%%%%%%%%%%%%%%%%%%%%%%%%%%%%%%%%%%%%%
%%%%%%%%%%%%%%%%%%%%%%%%%%%%%%%%%%%%%%%%%%%%%%
%\begin{center}
%\begin{table}[htbp]
\centering
\begin{tabular}{|c|c|c|c|}
\hline \hline
$(g_{B-L},~ M_{\widetilde{Z}_{B-L}})$  & ($0.1$,~ $5$ TeV)
                                & ($0.3$, ~ $5$ TeV)
                                & ($0.5$,~ $5$ TeV)     \\
\hline \hline
$m_{\tilde{\chi}^0_{1,2,3,4}}$           
                               & 132, 455, 728, 735
                               & 131, 455, 796, 802
                               & 130, 454, 832, 837 \\
$m_{\tilde{\chi}^{\pm}_{1,2}}$ 
                               & 133, 727
                               & 132, 794
                               & 131, 831 \\
$m_{\tilde{g}}$         
                         &  1297
                         &  1299
                         &  1300 \\
\hline
$m_{{\tilde{e},\tilde{\mu}}_{1,2}}$                   
                          & 606, 629 
                          & 1485, 1495
                          & 1752, 1761  \\
$m_{\tilde{\tau}_{1,2}}$ 
                          & 595, 626 
                          & 1477, 1491
                          & 1743, 1756  \\
\hline
$m_{{\tilde{u},\tilde{c}}_{1,2}}$  
                        
                         & 1228, 1240
                         & 1309, 1320
                         & 1346, 1357  \\
$m_{\tilde{t}_{1,2}}$   
                         & 990, 1132
                         & 1058, 1201
                         & 1084, 1230  \\
\hline
$m_{{\tilde{d},\tilde{s}}_{1,2}}$  
                        
                         & 1231, 1238
                         & 1312, 1319
                         & 1349, 1355  \\
$m_{\tilde{b}_{1,2}}$    
                         & 1099, 1223
                         & 1173, 1303 
                         & 1204, 1339  \\
\hline
$m_h$                    
                         & 124
                         & 125
                         & 126 \\
$m_H$                   
                         & 672
                         & 738
                         & 772 \\
$m_A$                   
                         & 672
                         & 737
                         & 772  \\
$m_{H^{\pm}}$           
                         & 677
                         & 742
                         & 776 \\
\hline \hline
\end{tabular}
%\caption{
%The same table as in Table~\ref{table1}.}
\caption{
Sparticle and Higgs boson mass spectra (in units of GeV) 
 in the case of $\tan \beta=10$, $F_\phi=50$ TeV and $f=2.5$.}
\label{table}
\vspace{-3cm}
\end{table}
%\end{center}
%%%%%%%%%%%%%%%%%%%%%%%%%%%%%%%%%%%%%%%%%%%%%%%%%%%

%%%%%%%%%%%%%%%%%%%%%%%%%%%%%%%%%
\end{document}